\documentclass[1p]{elsarticle}
\usepackage{graphicx}
\usepackage{amsmath}
\usepackage{amssymb}
\usepackage{slashed}

\newcommand{\be}{\begin{equation}}
\newcommand{\bpi}{\bar{\psi}}
\newcommand{\ee}{\end{equation}}

\newcommand{\eref}[1]{(\ref{#1})}
\newcommand{\fref}[1]{Fig.\ \ref{#1}}
\newcommand{\ie}{i.\,e.\ }
\newcommand{\gv}{g_{\textrm v}}
\newcommand{\gvc}{g_{\textrm v}^{\textrm{crit}}}
\newcommand{\Kc}{K_{\textrm c}}
 
\newcommand{\rhob}{\rho_\textrm{B}}

\begin{document}

\title{Role of Vector Interaction and Axial Anomaly in the PNJL Modeling of the QCD Phase Diagram}
\author{
Nino~Bratovic$^{1,2}$, Tetsuo~Hatsuda$^{2}$ and Wolfram~Weise$^{1}$\\
{\small ${^1}$Physik-Department, Technische Universit\"at M\"unchen, D-85747
Garching, Germany,}
\\{\small ${^2}$Theoretical Research Division, Nishina Center, RIKEN, Wako 351-0198, Japan}
}
\begin{abstract}
Effects of a flavor singlet vector interaction in the Polyakov - Nambu - Jona-Lasinio (PNJL) model are studied in combination with the axial U(1) breaking Kobayashi - Maskawa - 't Hooft interaction.
Using a consistent cutoff scheme we investigate the QCD phase diagram and its dependence on the vector coupling strength $\gv$. We find that the first order chiral phase transition at moderate baryon chemical potentials and its critical point, a generic feature of most NJL-type models without vector coupling, disappear for sufficiently large values of $\gv$ that are consistent with lattice QCD results at imaginary chemical potential. The influence of non-zero $\gv$ on the curvature of the crossover boundary in the $T-\mu$ plane close to $\mu = 0$ is also examined.
\end{abstract}

\maketitle

\section{{\it Introduction}}
The equation of state and the phase diagram of dense, strongly interacting matter, both at high and low temperatures, are in the focus of much recent interest, from heavy-ion collisions to the physics of compact stars. Lattice QCD computations continue to struggle with the treatment of (real) baryon chemical potentials $\mu$ when their ratios with respect to temperature T exceed  $\mu/T\sim1$. It is for this reason that investigations of the phase diagram at finite $\mu$ have so far relied mostly on models based on the symmetries and symmetry breaking patterns of low-energy QCD. 

Chiral models such as the Polyakov-loop-extended Nambu - Jona-Lasinio (PNJL) model \cite{Fu2003,Fukushima:2003fw,RTW2006,RRW2007,Hell2010} or the Polyakov-loop-improved quark-meson (PQM) model \cite{PQM,HPS2011} and most recently, Dyson-Schwinger equation approaches \cite{FLM2011}, 
are frequently used to calculate $(T,\mu)$ diagrams at large $\mu/T$.
In the region of interest, such models usually display a first order chiral phase transition (i.e.\ a pronounced discontinuity in the chiral condensate, $\langle \bar{q}q\rangle(\mu,T)$). More precisely, a chiral crossover starting from $\mu$ = 0 at the transition temperature $T_c \sim 0.2$ GeV proceeds at $\mu > 0$ until it reaches a critical $\mu_c$ at which the transition becomes first order. Starting from this critical point, a first-order transition line extends downward in temperature and meets the $\mu$ axis ($T=0$) typically at quark chemical potentials around $\mu \sim 0.3$ GeV, or baryon chemical potentials $\mu_B = 3\mu \sim 0.9$ GeV. The associated mixed phase covers a broad range of baryon densities $\rho_B$ from below $\rho_0 = 0.16$ fm$^{-3}$, the equilibrium density of nuclear matter, to about three times $\rho_0$. However, this is the nuclear physics domain where matter is known to be composed of interacting nucleons rather than PNJL type constituent quarks, and no chiral first-order phase transition is anywhere close. Hence the question must seriously be addressed \cite{ww2012} about the existence of a first-order phase transition (other than the nuclear liguid-gas transition) at low temperatures and $\mu_B\lesssim 1$ GeV, and consequently, about the existence of a critical point in the QCD phase diagram.   

The appearance of a first-order chiral phase transition in PNJL models is a characteristic feature of the simplest version of these models with a ``classic" NJL type interaction between quarks \cite{Nambu:1961tp}, i.e.\ a chiral combination of scalar and pseudoscalar interactions. It has already been noticed in previous work \cite{Asakawa:1989bq,Kitazawa:2002bc,Fukushima:2008wg,YTHB2007,ABHY2010} that the existence and location of the critical point in the phase diagram is extremely sensitive to additional interactions in the effective Lagrangian, especially to vector couplings and terms representing the axial $U(1)$ anomaly. In the present work, using a PNJL model, we systematically study the role of two interactions that are each capable of removing the first order transition completely from the phase diagram: 
a repulsive flavor singlet  vector interaction that shifts the baryon chemical potential, and the $\textrm{U}(1)_\textrm{A}$ breaking Kobayashi - Maskawa - 't Hooft interaction, the coupling strength of which also has a pronounced impact on the phase diagram. We focus in particular on the interplay of these two effects.

\section{\it The Model}
The present investigation is based on the three-flavor PNJL model \cite{Fu2003,Fukushima:2003fw,RTW2006,RRW2007,Hell2010} at mean field level. The NJL part of the Lagrangian for three quark flavors \cite{Hatsuda:1994pi,Vogl:1991qt,Lutz:1992dv,Rehberg:1995kh,Buballa:2003qv} has the general form:
\begin{eqnarray}
\mathcal{L}_{NJL}\,&=&\bar{\psi} \left(i\slashed{\partial}-\hat{m_0}\right)\psi+\frac{G}{2}\sum\limits_{a=0}^8\left[\left(\bar{\psi}\lambda^a\psi\right)^2+\left(\bar{\psi}i\gamma_5\lambda^a\psi\right)^2\right] \nonumber \\
&-&\frac{G_V}{2}\sum\limits_{a=0}^8\left[\left(\bar{\psi}\gamma_\mu\lambda^a\psi\right)^2+{G_A\over G_V}\left(\bar{\psi}\gamma_\mu\gamma_5\lambda^a\psi\right)^2\right] + \mathcal{L}_{\rm det}
\label{BasicLagrangian}
\end{eqnarray}
with the quark fields $\psi=\left(\psi_u,\psi_d,\psi_s\right)^\top$, the current quark matrix $\hat{m}_0=\textrm{diag}\left(m_u,m_d,m_s\right)$ and the Gell-Mann matrices $\lambda^a$ in $SU(3)$ flavor space. We work in the isospin-symmetric limit with $m_u = m_d$. For vanishing quark masses, the interaction terms in Eq.(\ref{BasicLagrangian}), apart from $\mathcal{L}_{\rm det}$, are invariant under the chiral $U(3)_L\times U(3)_R$ group. The $U(1)_A$ symmetry is broken by the axial anomaly in QCD. At the quark level, this symmetry breaking is introduced by adding the Kobayashi - Maskawa - t Hooft (KMT) interaction \cite{Kobayashi:1970ji,tHooft:1986}:
\be
\mathcal{L}_{\rm det}=K\left[{\rm det}_\textrm{f}\left(\bpi\left(1+\gamma_5\right)\psi\right)+{\rm det}_\textrm{f}\left(\bpi\left(1-\gamma_5\right)\psi)\right)\right],
\label{KMTHLagrangian}
\ee
with the KMT coupling strength $K$. Here $\textrm{det}_\textrm{f}$ are $3\times 3$ determinants in flavor space.
The combination of scalar-pseudoscalar, vector and axial vector interactions arises naturally if one starts from a 
QCD-inspired color current-current interaction and then performs a Fierz transform into color-singlet channels. In this case the coupling strengths have fixed values, $G_V = G_A = G/2$, relative to that of the scalar-pseudoscalar coupling $G$.

In the mean field approximation, dynamical quark masses are generated by the gap equations
\begin{align}
M_u&=m_u-\sigma_u+\frac{K}{2G^2}\sigma_u\sigma_s~,\\
M_s&=m_s-\sigma_s+\frac{K}{2G^2}\sigma_u^2~,
\label{EffectiveQuarkMasses}
\end{align}
with the chiral (light quark) condensate $\sigma_u = G\langle\bar{\psi_u}\psi_u\rangle = \sigma_d $ and the strange quark condensate $\sigma_s = G\langle\bar{\psi_s}\psi_s\rangle$. At non-zero quark densities $n_q = \langle\psi^\dagger\psi\rangle$, the flavor singlet term of the vector interaction, the one involving $\bar{\psi}\gamma_0\lambda^0\psi$, develops a non-zero expectation value while all other components of the vector and axial vector interactions have vanishing mean fields. In the present context we therefore focus on the reduced NJL Lagrangian
\begin{eqnarray}
\mathcal{L}_{NJL}\,=\bar{\psi} \left(i\slashed{\partial}-\hat{m_0}\right)\psi+\frac{G}{2}\sum\limits_{a=0}^8\left[\left(\bar{\psi}\lambda^a\psi\right)^2+\left(\bar{\psi}i\gamma_5\lambda^a\psi\right)^2\right] \nonumber + \mathcal{L}_{\textrm v}
 + \mathcal{L}_{\rm det}
%\label{BasicLagrangian}
\end{eqnarray}
and study the effects of the flavor singlet vector interaction 
\be
\mathcal{L}_{\textrm{v}}=-\frac{\gv}{2}\left(\bpi\gamma^{\mu}\psi\right)^2,
\label{VectorInteraction}
\ee
with varying coupling strength $\gv = {2\over 3}G_V$. Note that the presence of such an interaction with positive $\gv$ acts repulsively and shifts the quark chemical potential, $\mu\rightarrow\mu - \gv n_q$, so as to reduce the baryon density at given temperature $T$. We note at this point that a non-zero and positive vector coupling strength close to $\gv \simeq 0.4\, G$ was found to be important in reproducing two-flavor lattice QCD data for the phase diagram at imaginary chemical potential within the frame of a non-local PNJL model \cite{Hell:2011ic,Kashiwa:2011td}.

The $U(1)_A$ breaking KMT interaction is responsible for the large mass of the $\eta'$ meson relative to the remainder of the pseudoscalar meson octet. Indications of a possible in-medium $\eta'$ mass reduction \cite{Vertesi:2011xd} would suggests that $K$ might vary as a function of $T$ and $\mu$. We will treat $K$ as a constant but study results for different fractions $K/K_0\leq 1$, with $K_0$ being the vacuum value of the $U(1)_A$ breaking coupling strength. 

Amongst several similar NJL parameter sets available in the literature \cite{Lutz:1992dv,Hatsuda:1994pi,Rehberg:1995kh}, all adjusted to reproduce primarily physical masses and decay constants of the pseudoscalar meson octet, 
we use here the set from \cite{Lutz:1992dv}. We have checked that results with other sets differ in general only marginally.
The parameters determined by fitting meson properties in vacuum are $m_\textrm{u}=m_\textrm{d}=3.6$~MeV, $m_\textrm{s}=87$~MeV (consistent with current quark masses reported in \cite{pdg} at a renormalization scale of 2 GeV), a three-momentum cutoff $\Lambda=750$~MeV, $G\Lambda^2\approx3.6$ and $K_0\Lambda^5\approx8.9$.
The suggested value in \cite{Lutz:1992dv} for $\gv/G$ is 0.6 - 0.7, so that a meaningful window for
variations of the vector coupling is  $0.4 \lesssim \gv/G \lesssim 0.7$. 

The implementation of Polyakov loop dynamics in the PNJL model used here proceeds as in \cite{RRW2007}.
Quarks move via minimal gauge coupling in a homogeneous background color gauge field. 
The Polyakov loop itself is a path-ordered Wilson line winding around the imaginary time direction. The
expectation value $\Phi$ of the Polyakov loop, the order parameter for the confinement-deconfinement transition, is a real number, $\Phi=\Phi^*$, at mean field level. Its behavior is determined by the Polyakov loop effective potential,
\be
\mathcal{U}(\Phi,\Phi^*,T)=-\frac{1}{2}\left[a_0+a_1\left(\frac{T_0}{T}\right)+a_2\left(\frac{T_0}{T}\right)^2\right]\Phi^*\Phi+b_4\left(\frac{T_0}{T}\right)^3\log\left[ J\left(\Phi,\Phi^*\right)\right].
\label{TotalU}
\ee
with the Haar measure term
\be
J\left(\Phi,\Phi^*\right)= 1-6\Phi^*\Phi+4\left({\Phi^*}^3+\Phi^3\right)-3\left(\Phi^*\Phi\right)^2.
\ee
Parameters are determined by fits to pure gauge lattice data with $T_0 = 270$ MeV \cite{RRW2007}:
$a_0=3.51$, $a_1=-2.47$, $a_2=15.2$, $b_4=-1.75$. This
$\mathcal{U}$ is added to the fermionic thermodynamic potential.
The thermodynamic potential density in mean field approximation (with a zero-point energy contribution $\Omega_0$ subtracted) is then given by
\begin{eqnarray}
\Omega - \Omega_0 &=&
-T\sum\limits_j\int_{0}^{\Lambda}\frac{{\rm d}^3p}{\left(2\pi\right)^3}\,\ln\Bigl[1+e^{-E_j/T}\Bigl]-T\sum\limits_j\int_{\Lambda}^{\infty}\frac{{\rm d}^3p}{\left(2\pi\right)^3}\,\ln\Bigl[1+ e^{-E^{\rm free}_j/T}\Bigl]\notag\\
&+& ~\mathcal{U}\left(T,\Phi,\Phi^{*}\right)~ +~ \frac{\sigma_u^2}{2G} ~+ ~\frac{\sigma_s^2}{4G}~ -~ \frac{K}{2G^3}\sigma^2_u\sigma_s - {\gv\over 2} n_q^2~.
\label{ThDynPot}
\end{eqnarray}
The $E_j(|\vec{p}\, |;\sigma, \mu, \Phi)$ are the quasiparticle energy eigenvalues from the fermionic propagator matrix
and $\mathcal{U}$ the Polyakov loop effective potential detailed above in \eref{TotalU}.
The regularization of $\Omega$ with a three-momentum cutoff $\Lambda$ is motivated as follows.
The first term on the r.h.s. is the thermal quark quasiparticle energy contribution from momentum regimes below the cutoff where the NJL interaction is ``turned on". There the quasiparticles propagate in a condensate background
with dynamically generated effective quark masses given in \eref{EffectiveQuarkMasses}.
For momenta above the cutoff $\Lambda$, the NJL interaction is ``off", and the quarks propagate as free particles with their current quark masses under the sole influence of the Polyakov loop.
In this high-momentum regime all couplings are set to zero in accordance with the basic premises of the NJL model, \ie
\begin{equation}
 E_j\to E_j^\textrm{free} = \sqrt{\vec{p}\,^2 + m_j^2} \pm\mu\,\,; \qquad G\ =\ K\ =\ \gv\ =0 \qquad \textrm{for}\qquad |\vec{p}\,|>\Lambda
\label{SetCouplingsZero}
\end{equation}
(with $-\mu$ for quarks, $+\mu$ for antiquarks).
This procedure avoids unphysical behavior of the quark condensates and guarantees that the pressure reaches the Stefan--Boltzmann limit at high temperatures.  
Indeed, the chiral condensates vanish in the high temperature limit so that the quarks are free and carry just their current masses according to Eq.\eref{EffectiveQuarkMasses}. \footnote{The nonlocal version of the PNJL model \cite{Hell2010,Hell:2011ic} is a further improvement that avoids the momentum cutoff altogether by introducing a momentum-dependent dynamical quark mass $M(p)$ that connects smoothly to the perturbative QCD limit as $p\rightarrow\infty$.} 

Finally, the mean fields entering $\Omega$ are determined by the self-consistent equations
\begin{equation}
\frac{\partial\,\,\Omega}{\partial \sigma_u}= \frac{\partial\,\,\Omega}{\partial \sigma_s}= \frac{\partial\,\,\Omega}{\partial \Phi}=0
\label{SelfConsistencyEquations}
\end{equation}
and
\be
-\frac{\partial\,\Omega}{\partial \mu}=n_q = 3\,\rho_B
\ee
with the quark baryon number density, $n_q$, and the baryon density, $\rho_B$.

\section{\it Results}

It has already been known that a term like $\eref{VectorInteraction}$ lowers the position in temperature of the critical point with increasing $\gv>0$, until it disappears completely from the phase diagram at a sufficiently large $\gv$ \cite{Asakawa:1989bq,Fukushima:2008wg,Carignano:2010ac}. In our case, this happens for $\gv\ge\gvc\simeq 0.38\,G$.
Lowering $K$ leads to a similar effect \cite{Fukushima:2008wg}. With the parameter set used here, the critical point disappears for $K \le\Kc\simeq 0.71\,K_0$.

\begin{figure}[htb]
\begin{minipage}[t]{6.5cm}
\includegraphics[width=6cm]{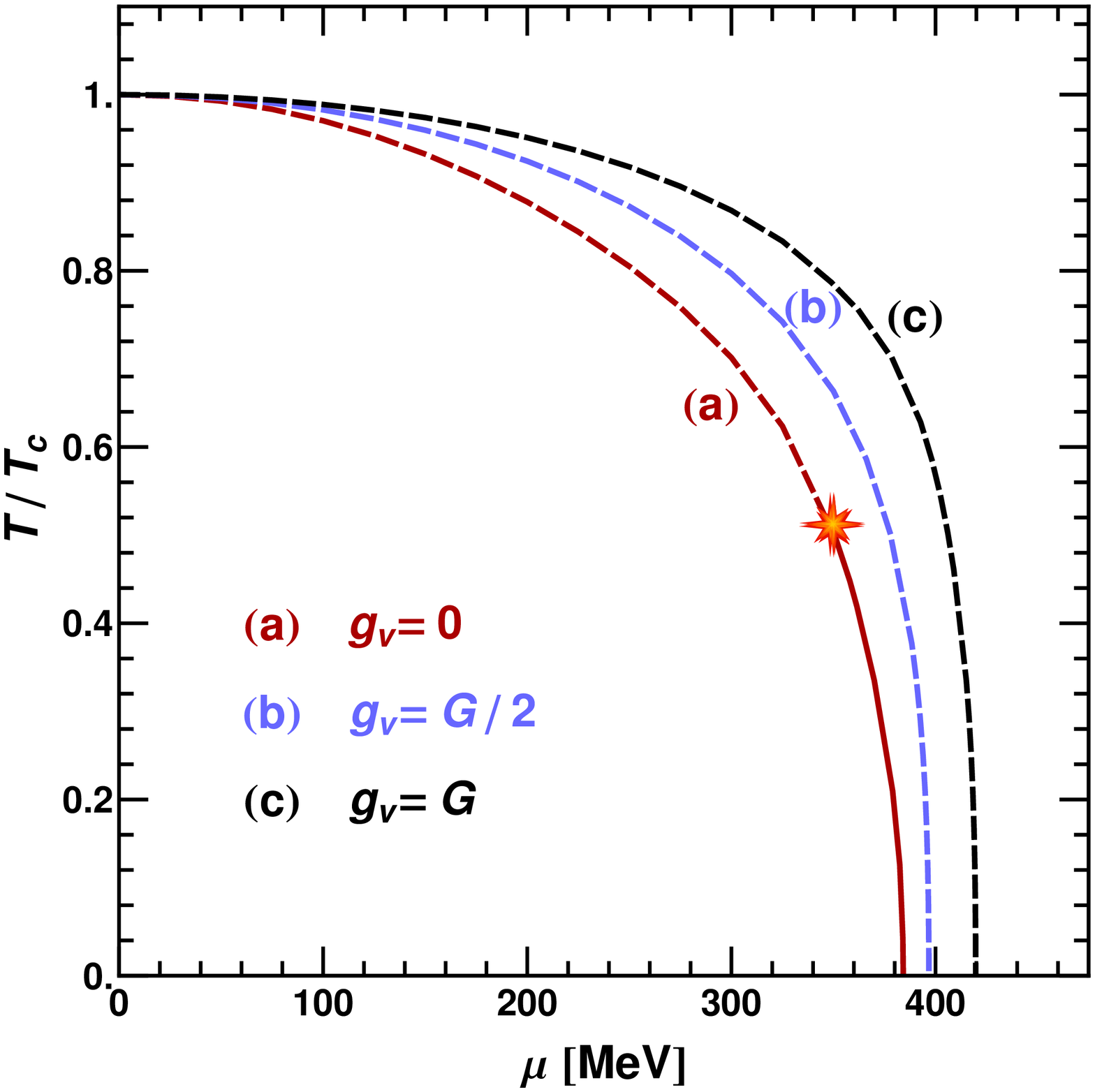}
\caption{PNJL phase diagrams for three different vector coupling strengths:
(a) $\gv=0$, (b) $\gv=G/2$ and (c) $\gv=G$.
Dashed lines denote $\sigma_u\left(T,\mu\right)/\sigma_u\left(T=0,\mu=0\right)=0.5$ at crossover transitions.
The solid line is the first order transition ending in the critical point.} 
\label{fig1}
\end{minipage}
\hspace{\fill}
\begin{minipage}[t]{6.5cm}
\includegraphics[width=6.3cm]{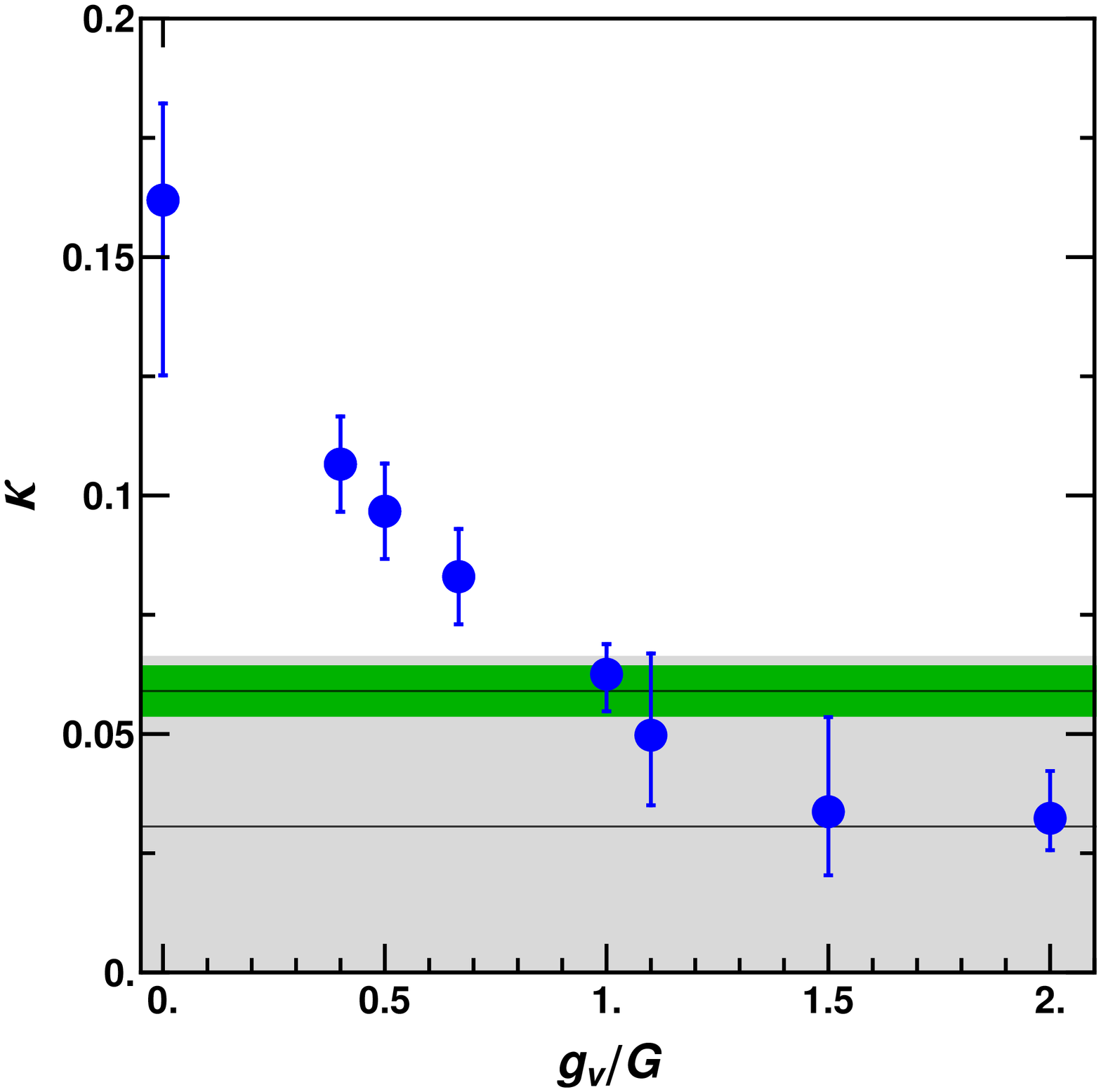}
\caption{Curvature $\kappa$ of the crossover boundary near $T_c$ as function of the vector
coupling strength $\gv$. Dots: results of calculations using the same input as in \fref{fig1}.
Uncertainty measures attached to the dots give impression of the half width over which the crossover spreads.  Upper horizontal line with narrow error band: (2+1)-flavor lattice QCD result of Ref.\cite{Kaczmarek:2011}; 
broad grey band: result of Ref.\cite{Endrodi:2008}.} 
\label{fig2}
\end{minipage}
%\hspace{\fill}
\end{figure}

We have studied the effects of variations of both parameters, $\gv$ and $K$, on the $T$--$\mu$ and $T$--$\rhob$ phase diagrams. First, consider the phase diagram in the $T$--$\mu$ plane for three choices of $\gv$ in \fref{fig1}. The case $\gv = 0$ displays the familiar picture of a chiral 
crossover terminating as a second order transition at a critical point. From this critical point downward a first order chiral phase transition extends to the $\mu$ axis at $T = 0$. For 
$\gv = G/2$ the first order phase transition has already disappeared and turned into a continuous crossover all the way down in temperature. For even larger vector coupling, $\gv \simeq G$, the crossover pattern is shifted to higher chemical potentials. Note that what is
represented here as a ``crossover line" does not mean that chiral symmetry is restored in its Wigner-Weyl realization
beyond that line. It just means that, along the line, the chiral condensate is reduced in magnitude to half of its value at zero temperature and $\mu = 0$. The crossover is in fact smooth and extends to densities more than four times the baryon density of normal nuclear matter.

%figure-----------------------------------------------------------------------------------------------------------------------------
%
%
%figure-----------------------------------------------------------------------------------------------------------------------------

Shifting the chiral crossover to regions of larger $\mu$ implies a flattening of the curvature 
of the crossover boundary in the neighborhood of $\mu = 0$. This is shown in \fref{fig2}. The curvature
of the crossover boundary is defined by
\begin{equation}
{T_c(\mu)\over T_c(0)} = 1 - \kappa\left({\mu^2\over T_c^2}\right) , ~~~ \textrm{or~equivalently,}~~~\kappa = - T_c{dT_c(\mu)\over d\mu^2}\Big|_{\mu^2 = 0}~.
\end{equation}
Standard NJL or PNJL calculations without vector interaction usually produce a curvature parameter $\kappa$ that is too large in comparison with lattice results. A recent accurate lattice QCD computation \cite{Kaczmarek:2011} gives
$\kappa \simeq 0.06$. Evidently, a reasonably large vector coupling strength $\gv$ is capable of approaching such a small  curvature.

%figure-----------------------------------------------------------------------------------------------------------------------------
%
\begin{figure}[htb]
\begin{center}
\includegraphics[width=6cm]{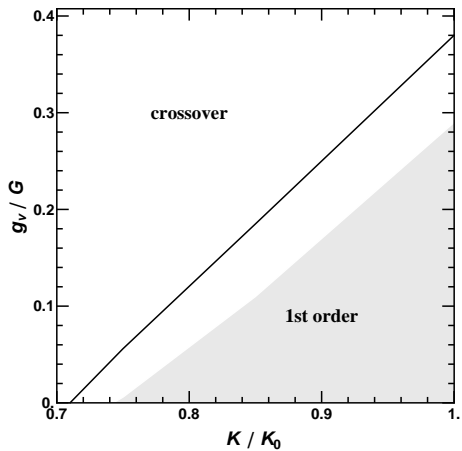}
\caption{Phase structure in the PNJL model displayed as a function of the vector coupling strength $\gv$ (relative to the scalar-pseudoscalar coupling G) and the $U(1)_A$ breaking (KMT) coupling $K$ (relative to its vacuum value
$K_0$). The diagonal line separates regions in which a first order chiral phase transition takes place from those with smooth crossover. The grey area indicates the (unphysical) domain in which the mixed-phase region of a chiral first-order transition would overlap with the coexistence region of the liquid-gas phase transition in nuclear matter as calculated using in-medium chiral effective field theory \cite{Fiorilla:2011sr}.} 
\end{center}
\label{fig3}
%\hspace{\fill}
\end{figure}
%
%figure-----------------------------------------------------------------------------------------------------------------------------

Next, consider the phase structure in scenarios with varying vector coupling $\gv$ and KMT coupling $K$.
Increasing $\gv$ as well as reducing $K$ has the effect of removing the first order chiral phase transition
altogether from the phase diagram. The systematics is displayed in \fref{fig3}. Evidently, the appearance or disappearance of a first order phase transition is very sensitive to both the vector interaction and the strength of the axial $U(1)$ anomaly. For any value $K \leq K_0$, the 1st order transition turns into crossover when $\gv\gtrsim 0.4\,G$. If $K$ is reduced in the medium relative  to the vacuum, the crossover scenario is initiated for even smaller values of $\gv/G$. (We recall again that preferred values of $\gv$ range between 0.4 and 0.7 judging from lattice QCD thermodynamics at imaginary chemical potential and from earlier NJL phenomenology.)

The first order chiral transition realized at $\gv=0$ and viewed in the plane of temperature and baryon number density $\rhob$ is known to run into an unphysical situation \cite{ww2012}: the phase mixing region associated with this 1st order chiral transition overlaps with the coexistence region of the liquid-gas phase transition in nuclear matter, 
realistically calculated within the framework of in-medium chiral effective field theory \cite{Fiorilla:2011sr} and properly using nucleons rather than quarks as active fermionic degrees of freedom.  Such an overlap, marked by the grey area in \fref{fig3}, is simply forbidden by well known facts of nuclear physics. This unphysical behavior is removed as soon as the repulsive vector interaction is introduced with reasonable values of its coupling strength, $\gv\gtrsim 0.4\,G$, and the chiral transition turns to a smooth crossover.

\section{\it Summary and Discussion}

1. A first lesson from the present study is the strong sensitivity of the existence and location of a critical point to vector and $U(1)_A$ breaking interactions in any NJL or PNJL model. Realistic magnitudes of the vector coupling strength
avoid the first-order chiral phase transition altogether and likewise dismiss the associated existence of a critical point in the phase diagram.

2. While the exact value of the vector coupling strength is uncertain, it is likely to be larger than 0.4 times the 
strength of the chiral scalar-pseudoscalar interaction in NJL type models. This is sufficient to induce a smooth
chiral crossover scenario at high densities and low temperatures, leaving well known nuclear physics constraints
at lower baryon densities intact.

3. A sufficiently strong flavor singlet vector interaction of the kind discussed in this work is likely to provide the necessary repulsion at high density that helps supporting two-solar-mass neutron stars \cite{Demorest:2010bx} by making the equation of state of such systems sufficiently stiff \cite{Ozel:2010fw,Lattimer:2010uk}. Work along these lines is in progress.

In summary, considering results from various sources using diverse methods,
we tend to find a strong preference for a QCD phase diagram without a first order chiral phase transition, but rather
with a smooth crossover at high densities and low temperature. \\

{\it Acknowledgements}. This work was supported in part by BMBF, GSI, the DFG Excellence Cluster ``Origin and Structure of the Universe", and by JSPS grant No.\ 22340052. Useful and stimulating discussions with Kenji Fukushima,  Thomas Hell and Jan Pawlowski are gratefully acknowledged. Two of us (N.B. and W.W.) thank the RIKEN Nishina Center for kind hospitality; N.B. thanks for support by the RIKEN IPA program and the TUM Graduate School.

\bibliographystyle{srt}

\end{document}